\documentstyle[preprint,aps]{revtex}
\input epsf.tex
\def\DESepsf(#1 width #2){\epsfxsize=#2 \epsfbox{#1}}
\begin{document}
\preprint{\vbox{\hbox{OITS-630}\hbox{}}} 
\draft
\title{Contact Interaction Explanation of HERA Events\\ and $SU(3)_C\times
SU(2)_L\times U(1)_Y$ Invariance}
\author{N.G. Deshpande$^1$, B. Dutta$^1$ and Xiao-Gang He$^2$}
\address{$^1$Institute of Theoretical Science, University of Oregon\\ Eugene, OR
97403-5203, USA\\ and\\
$^2$School of Physics, University of Melbourne\\ Parkville, Vic. 3052, Australia}
\date{May, 1997}
\maketitle
\begin{abstract} We reexamine the contact interaction hypothesis for
excess high-$Q^2$ events seen at  HERA in $e^+ p \rightarrow e^+ X$ scattering.
We consider the most general  structure of contact interactions  insisting on
$SU(3)_C\times SU(2)_L
\times U(1)_Y$ invariance. Apart from constraints for $e^+ e^-\rightarrow q\bar
q$ , Drell-Yan and atomic parity violation experiments, we find stringent
constraints from neutrino deep inelastic data, provided one assumes lepton
universality. A possible choice of contact
terms that escape present bounds is still possible, althought new data from
FERMILAB and LEP should further constrain such a possibility.
\end{abstract}
\pacs{PACS numbers:}

\newpage

Many explanations of the excess in  $e^+p \rightarrow e^+ X$ scattering events 
observed at large $Q^2$ at HERA\cite{hera-ex} have been offered in the
literature\cite{hera-th}. These explanations either call for one or more s
channel leptoquark  added to the Standard Model (SM) or embedded within
supersymmetric framework, or call for contact interactions that may arise due to
an  exchange of heavy particles beyond the SM. A statistical fluctuation from
the SM can also not be ruled out.

In this note we shall pursue the hypothesis that the excess of events arises 
due to contact interactions. These interactions can come from exchange of
particles  in s or t channel, and we consider the most general structure. Any
physics beyond the SM must necessarily respect the low  energy $SU(3)_C\times
SU(2)_L\times U(1)_Y$ symmetry. We thus impose this symmetry on all  the new
terms. This symmetry along with the lepton universality allows us
to use neutrino data to arrive at more stringent limit than those considered
previously \cite{barger,ellis}. Reference \cite{bf} also invokes this
symmetry but does not strictly impose the constraints on the solutions. 

Under the SM gauge group, the left-handed ($Q_L$),  right-handed ($U_R$, $D_R$) 
quarks and left-handed ($L_L$), right-handed ($E_R$) leptons transform as:
\begin{eqnarray} Q_L &:& (3,\;2,\;1/3)\;; U_R : (3,\;1,\;4/3)\;; D_R :
(3,1,-2/3)\;;\nonumber\\ L_L &:& (1,\;2,\;-1)\;; E_R : (1,1,-2)\;.
\end{eqnarray}

Contact interactions relevant to HERA data must contain quarks and leptons. The
following is a list of the allowed quark-lepton contact interactions invariant
under the SM gauge symmetry\cite{contact}:

\begin{eqnarray} O_{LL} &=& \bar L_L \gamma^\mu L_L\bar Q_L \gamma_\mu Q_L \;;
O_{RR}^u = \bar e_R \gamma^\mu e_R\bar u_R \gamma_\mu u_R \;;\nonumber\\
O_{RR}^d &=& \bar e_R \gamma^\mu e_R\bar d_R \gamma_\mu d_R \;; O_{RL} = \bar
e_R \gamma^\mu e_R\bar Q_L\gamma_\mu Q_L \;;
\nonumber\\
 O_{LR}^u &=& \bar L_L\gamma^\mu L_L\bar u_R\gamma^\mu u_R\;;
 O_{LR}^d = \bar L_L\gamma^\mu L_L\bar d_R \gamma_\mu d_R \;;
\nonumber\\ O_V &=& \bar Q_L\gamma_\mu L_L \bar e_R \gamma^\mu d_R\;; O_S = \bar
Q_L e_R \bar L_L u_R\;;\nonumber\\ O_{LL}^3 &=& \bar L_L^i\gamma^\mu L_L^k\bar
Q_L^j \gamma_\mu Q_L^l
\epsilon_{ij}\epsilon_{kl} \;.
\end{eqnarray} and  parameterize the effective
Lagrangian added to the SM Lagrangian as
\begin{eqnarray} L_{contact} = \sum_i \eta_i O_i\;.
\end{eqnarray} Here $i$ runs the same indices as the indices of the operators in
equation (2). The parameters $\eta_i$ represent the strength for the new
interactions. Assuming CP conservation in these contact interactions, $\eta_i$
are real and can have positive or negative signs. We are mostly concerned about
the first generation contact terms in (3), however in order to use CCFR limits,
we shall make the assumtion of lepton universality.

There are many possible mechanisms which can generate the above listed contact
interactions. Exchanges of super heavy particles such as $Z'$, vector
leptoquark,  scalar leptoquark are some of the possibilities. 

Let us now analyse which operators that are relevant to HERA data. We first
consider operators
$O_{V,S}$. The strength of these two  operators are severly constrained from low
energy data on $\pi^-\rightarrow  e\bar \nu$ because they lead to enhanced
matrix elements compared with the SM contribution. A simple estimate gives
\begin{eqnarray} <e\nu|O_V|\pi^-> &=& -2 \bar e_R \nu_L <0|\bar u_L d_R|\pi^->
=i\sqrt{2} f_\pi {m_\pi^2\over m_u+m_d} \bar e_R \nu_L\;,\nonumber\\
<e\nu|O_S|\pi^-> &=& -{1\over 4} <e\nu|O_V|\pi^->\;,
\end{eqnarray} where $f_\pi = 93$ MeV is the $\pi$ decay constant.

It has been shown that the SM prediction for $\pi^-\rightarrow e\bar \nu$ 
agrees with the experimental data within 1\%\cite{pion}. Requiring that new
physics does not change the SM result by more than 1\%, we find that
$\eta_V$ and $\eta_{S}$ are constrained to be less than $0.5\times 10^{-4}$
(TeV$^{-2}$) and 
$ 2\times 10^{-4}$ (TeV$^{-2}$), respectively. With these bounds, the effects of
$O_{V}$ and $O_S$ are too  small to have any impact on HERA data.

The remaining contact interactions  are in fact of the same type as analysed in
Ref.\cite{barger} for $e \bar e q \bar q$ contact  interactions. However,
because the requirement that the contact interactions be invariant under
$SU(3)_C\times SU(2)_L\times U(1)_Y$, there are several fundamental differences. 
We note that $\eta_{RL}^{eu} = \eta_{RL}^{ed} = \eta_{RL}$, and for left-handed
electron in the
$e \bar e q \bar q$ interactions, there are also associated $\nu \bar \nu q \bar
q$ interactions. These constrain the interactions further.

We now analyse the high-$Q^2$ $e^+p$ events at HERA.  The cross section for $e^+
p \rightarrow e^+ X$ is give by
\begin{eqnarray} {d\sigma(e^+p)\over dx dy} &=& {sx\over 16 \pi} 
\{u(x,Q^2)[|M_{LR}^{eu}|^2 + |M_{RL}^{eu}|^2 +(1-y)^2(|M_{LL}^{eu}|^2 +
|M_{RR}^{eu}|^2)]\nonumber\\ &+&d(x,Q^2)[|M_{LR}^{ed}|^2 + |M_{RL}^{ed}|^2
+(1-y)^2(|M_{LL}^{ed}|^2 + |M_{RR}^{ed}|^2)]\}\;,
\end{eqnarray} where $Q$ is the momentum transfer, 
$u(x,Q^2)$ and $d(x,Q^2)$ are the u and d quark parton distributions, and
\begin{eqnarray} x &=& {Q^2\over 2 P.q}\;,\;\; y = {Q^2\over s x}\;,\nonumber\\
M_{ij}^{lq} &=& -{e^2Q_eQ_q\over sxy} - {g_Z^2(T^3_{ei} -
\mbox{sin}^2\theta_WQ_e)(T^3_{qj}-\mbox{sin}^2\theta_WQ_q)\over sxy+m_Z^2} +
\eta_{ij}^{eq}\;.
\end{eqnarray} Here $Q_f$ and $T^3_{fi}$ are charges and weak isospin,
respectively, $g_Z = e/(\mbox{sin}\theta_W
\mbox{cos}\theta_W)$,  $P$ and $q$ are the incoming proton and  positron
momentum, respectively, and $s$ is the CM energy-squred.

Using CTEQ-3 parton distribution functions, we find that to explain the excess
of events at HERA, the size of the contact terms necessary taking them singly,
is $\eta_{LR}^u$, $\eta_{RL}$ approximately 1.5 TeV$^{-2}$, $\eta_{LR}^d$
approximately 6 TeV$^{-2}$. Contact terms of the type $\eta_{LL}^3$,
$\eta_{LL}$, $\eta_{RR}$ have to be much larger, approximately 8 TeV$^{-2}$
because these contributions are suppressed by the factor $(1-y)^2$ and favor
excess events at low y. Experimental data for the y distribution of the  excess
events clearly favors excess events at  high y\cite{hera-ex}.

Possible solutions to HERA events must satisfy constraints from other
experimental data.  We now consider the limits on the contact interactions from
$e^+e^-$ collider data at LEP.  Table 1 gives the OPAL preliminary 95\%
C.L.\cite{opal} limits on various contact terms converted to our notation.
Similarly in Table 2 we give the CCFR limits\cite{ccfr} remebering that the data
was from neutrino on iron  target, which has roughly equal numbers of neutrons
and protons. The limits on $\eta_{LL}$, $\eta_{LL}^3$, $\eta_{LR}^u$ and
$\eta_{LR}^d$ are much stronger from CCFR than those from OPAL.

Finally we have extremely strong limits from atomic parity violation parameter
$Q_W$.  For 
$^{133}_{55}Cs$, we have for change in $Q_W$\cite{barger} from the SM:
\begin{eqnarray}
\Delta Q_W &=& (11.4\mbox{TeV}^2) (\eta_{LL}^3+ \eta_{LL} + \eta_{LR}^u -
\eta_{RL} -\eta_{RR}^u)\nonumber\\ &+&(12.8\mbox{TeV}^2) (\eta_{LL} +\eta_{LR}^d
- \eta_{RL} -\eta_{RR}^d)\;.
\end{eqnarray} Experimental measurements find $Q_W = -72.11\pm0.93$\cite{atomic}
while the SM prediction for 
$m_t = 175$ GeV and $m_H = 100$ GeV is $Q_{W}^{SM} = -73.04$. It is clear that
the difference 
$\Delta Q_W = 1.09 \pm 0.93$ places a severe constraint on the allowable contact
interactions, and it can not accommodate any single contact term. 

It is clear that any solutions with $O_{LL}$ or $O_{LL}^3$ term giving
significant contribution to the HERA events are ruled out because large
$\eta_{LL}$ or $\eta_{LL}^3$ required are in conflict with OPAL and CCFR data.
It is however possible to have solutions with significant contribution from
$O_{LR}$ or $O_{RL}$. Given that we need significant $\eta_{LR}$ or $\eta_{RL}$
couplings to explain HERA data, we can try to obtain solutions that satisfy all
the constraints. The following are two types of solutions:

{\bf Solution A}

\begin{eqnarray}
\eta_{LR}^u &=& \eta_{RL} = \eta_{LR}^d\;,\nonumber\\
\eta_{LL} &=& \eta_{LL}^3 = \eta_{RR}^u = \eta_{RR}^d = 0\;.
\end{eqnarray}

This solution corresponds to the one first suggested by  Ref.\cite{barger}
and also discussed in Ref.\cite{bf}. However
note that we must include both u and d  quark contributions because of gauge
invariance requirement and we are restricted to having
$\eta_{LR}^u$ and $\eta_{LR}^d$ $<1.01$ TeV$^{-2}$ from CCFR data. This solution
is barely able to explain the excess events. We note that Ref.\cite{barger}
preferred a values of $\eta_{LR}^u \approx 1.4$ TeV$^{-2}$. Analysis of new data
from FERMILAB is expected to improve this bound significantly making this
solution probably inadmissable.

{\bf Solution B}

\begin{eqnarray}
\eta_{RL} &=& -\eta_{RR}^u = -\eta_{RR}^d\;,\nonumber\\
\eta_{LR}^u &=& \eta_{LR}^d = \eta_{LL} =\eta_{LL}^3 = 0\;.
\end{eqnarray}

The most strigent limit now comes from OPAL and Drell-Yan process. However,
$\eta_{RL} \approx 1.5
$ TeV$^{-2}$ is clearly acceptable. In Figure 1 we plot $\sigma(Q^2>Q^2_{min})$
as a  function of $Q^2_{min}$ for this  and compare it with SM and data. Since
we have $\eta_{RR}^u$ and $\eta_{RR}^d$ terms, the y distribution is now
altered. However, the change is not significant and this choice is consistant
with data. To demonstrate this, in Figure 2 we show the y distribution of events
expected with addition of contact terms and compare it with the SM.  In Figure 3
we show the Drell-Yan process with the contact term and compare it to the SM. In
the region of invariant dilepton mass up to 300 GeV there is little deviation
from the SM, and CDF data is in excellent agreement \cite{bodek}. Beyond that,
between 400 GeV to 800 GeV, contact interactions produce an excess of events
which  for integrated luminosity of 110 $pb^{-1}$ are approximately 10 compared
to 1 in SM. Data at present is consistent with SM, though one can not rule out
contact interaction. Further gain in luminosity will constrain this model,
making this explanation of HERA data unacceptable.

In the above, we have assumed that the contact interactions only involve the
first generation of quarks. In general the contact interacitons will
also involve other generations. If these interactions are due to leptoquark
exchange, the couplings involving different generations can have different
strength. It is possible  that couplings to other generations are small and will
not cause any difficulties. However, if the contact interactions are due to
exchange of $Z'$ particles \cite{sg} with universal coupling to different 
generations (leptons and quarks), the
contact interaction explanation of the excess events at HERA could be in trouble
because the constraints from various  experiments become even tighter. 

For example,  a class of $Z'$ models inspired by string  thoeries based on $E_6$
even without coupling constant constraint from GUT is ruled out. These models
have charges given by\cite{gut} $Q(\alpha) = Q_\psi \mbox{cos}(\alpha) + 
Q_{\chi} \mbox{sin}(\alpha)$, where $Q_\psi$ and $Q_{\chi}$ are the charges of
$U(1)_\psi$ and $U(1)_\chi$ subgroup of $E_6$ symmetry. The contact interaction
generated by this  model is
\begin{eqnarray}
\eta_{ij}^{lq} &=& \eta_0 Q^l_i(\alpha) Q^q_j(\alpha)\;,\nonumber\\
Q^e_L(\alpha) &=& -Q^d_R(\alpha) = Q^\nu_L(\alpha) = \mbox{cos}(\alpha) +
\mbox{sin}(\alpha)
\sqrt{{3\over 5}}\;,\nonumber\\ Q^e_R(\alpha) &=& -Q^u_L(\alpha) = Q^u_R(\alpha)
= -Q^d_L(\alpha) = -[\mbox{cos}(\alpha)  -3\sqrt{{3\over
5}}\mbox{sin}(\alpha)]\;,
\end{eqnarray} and $\eta_0$ is an arbitrary free parameter since we do not
impose constraints from coupling unification. Constraint from $Q_W$ now yields,
\begin{eqnarray}
\Delta Q_W = \eta_0 (102.4)\mbox{TeV}^{2} [\sqrt{{3\over 5}}\mbox{cos}\alpha\,
\mbox{sin}
\alpha - {3\over 5} \mbox{sin}^2\alpha]\;.
\end{eqnarray} $\eta_0$ has to be of order 1 TeV$^{-2}$ to explain HERA
data and the only acceptable  solutions are
$\mbox{tan}\alpha = 0$ and $\sqrt{5/3}$. By considering $\eta_{LL}$, neutrino
deep inelastic  scattering data now yields a bound on $\eta_0$ to be less than 
$ 0.5 $ TeV$^{-2}$ for $\alpha = 0$ and 
$0.3$ TeV$^{-2}$ for
$\mbox{tan}\alpha = \sqrt{5/3}$. Further, LEP $e^+e^-$ data imposes strong
constraint on 
$\eta_0$ for $\mbox{tan}\alpha = \sqrt{5/3}$ solution,  requiring $\eta_0 <
0.15$ TeV$^{-2}$ from the requirement that cross-section at large $s$  shall not
deviate  from SM by more than 6\%.

In conclusion, we have examined the contact interaction explanation for
high-$Q^2$ events  seen at  HERA in $e^+ p \rightarrow e^+ X$ scattering. We
considered the most general  structure of contact interactions  invariant under
$SU(3)_C\times SU(2)_L
\times U(1)_Y$. Experimental data from $e^+e^-\rightarrow q\bar q$, Drell-Yan, 
deep inelastic neutrino-nucleon processes, and atomic parity violating
interactions place  severe bounds on the strengths of contact interactions.
Assuming the contact interactions only  involve the first generation of leptons
and quarks, we found that there are solutions which escape present bounds
although new data  from FERMILAB and LEP should further constrain such
possibility.  Contact interactions generated by $Z'$ exchange are more
stringently constrained.

We would like to thank V. Barger, K. Cheung, R. Frey and D. Strom  for useful 
discussions.
We would also like to thank Cao Zhen for helping us with a new Garphics
software and M. Mangano for useful communication. 
This work was  partly supported  by a DOE  grant no.
DE-FG06-854ER-40224. XGH was supported by Australian Research Council.

\newpage
\begin{center}  Table 1 \end{center}

\begin{center}
\begin{tabular}{|c|c|c|}\hline &$e^+e^-\rightarrow u \bar u$&$e^+e^-\rightarrow
d
\bar d$\\&+\hspace{0.5cm}-&+\hspace{0.5cm}-\\\hline
$\eta_{LL}$&10.4\hspace{0.5cm}2.2&2.2\hspace{0.5cm}12.6\\
$\eta_{LL}^3$&10.4\hspace{0.5cm}2.2&\\
$\eta_{RR}^u$&6.4\hspace{0.5cm}4.4&\hspace{0.5cm}\\
$\eta_{RR}^d$&\hspace{0.5cm}&2.9\hspace{0.5cm}8.7\\
$\eta_{RL}$&4.4\hspace{0.5cm}6.4&4.9\hspace{0.5cm}5.6\\
$\eta_{LR}^u$&5.6\hspace{0.5cm}4.9&\hspace{0.5cm}\\
$\eta_{LR}^d$&\hspace{0.5cm}&4.4\hspace{0.5cm}6.4\\\hline
\end{tabular}
\end{center}
\noindent{\bf Table Caption:} OPAL 95 $\%$ C.L. limits on contact interactions.
 The signs + and - indicate the signs of $\eta_i$.
\vspace{2 cm}

\begin{center}  Table 2 \end{center}

\begin{center}
\begin{tabular}{|c|c|c|}\hline &$+$&$-$\\\hline
$\eta_{LL}$&0.57&0.48\\
$\eta_{LL}^3$&0.80&0.68\\
$\eta_{LR}^u$&1.01&0.92\\
$\eta_{LR}^d$&1.01&0.92\\\hline
\end{tabular}
\end{center}
\noindent{\bf Table Caption:}CCFR 95$\%$ C.L. limits on contact interactions.
The signs + and -
indicate the signs of $\eta_i$
\newpage
\leftline{{\Large\bf Figure captions}}
\begin{itemize}
\item[Fig. 1~:] {Cross-section($Q^2>Q^2_{\rm min}$ is shown against minimum
$Q^2$. The solid line corresponds to solution B and the dotted line corresponds
to SM. The data points (combined H1 and ZEUS measurements) are shown for
$Q^2\ge15000 GeV^2.$ }.
\item[Fig. 2~:] {y  distribution is shown for $Q^2>15000 GeV^2$ for
SM(dotted) and for solution B(solid) }

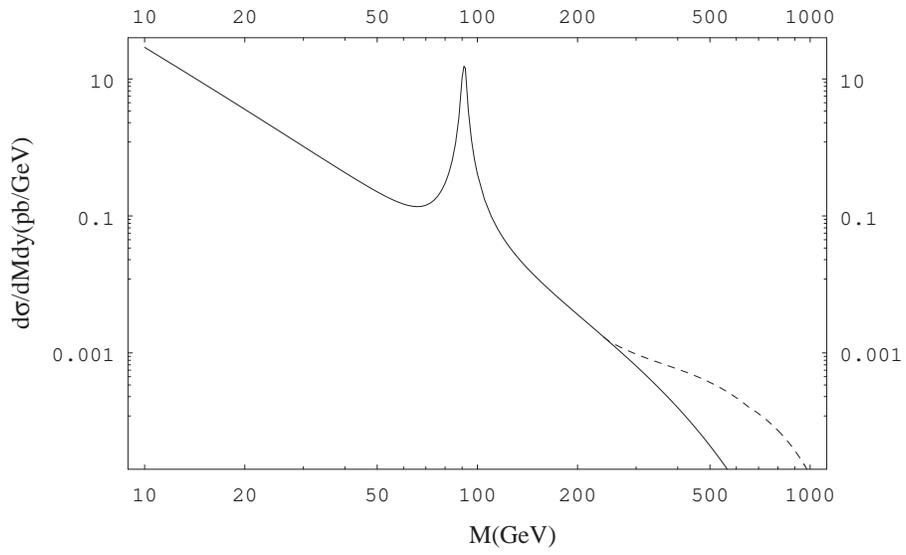
\item[Fig. 3:] {The Drell-Yan cross-section (for $\left |y \right|<1$,y is the
rapidity) at the Tevatron is shown. The solid line corresponds to SM and the
dotted line corresponds to the solution B. }
\end{itemize}
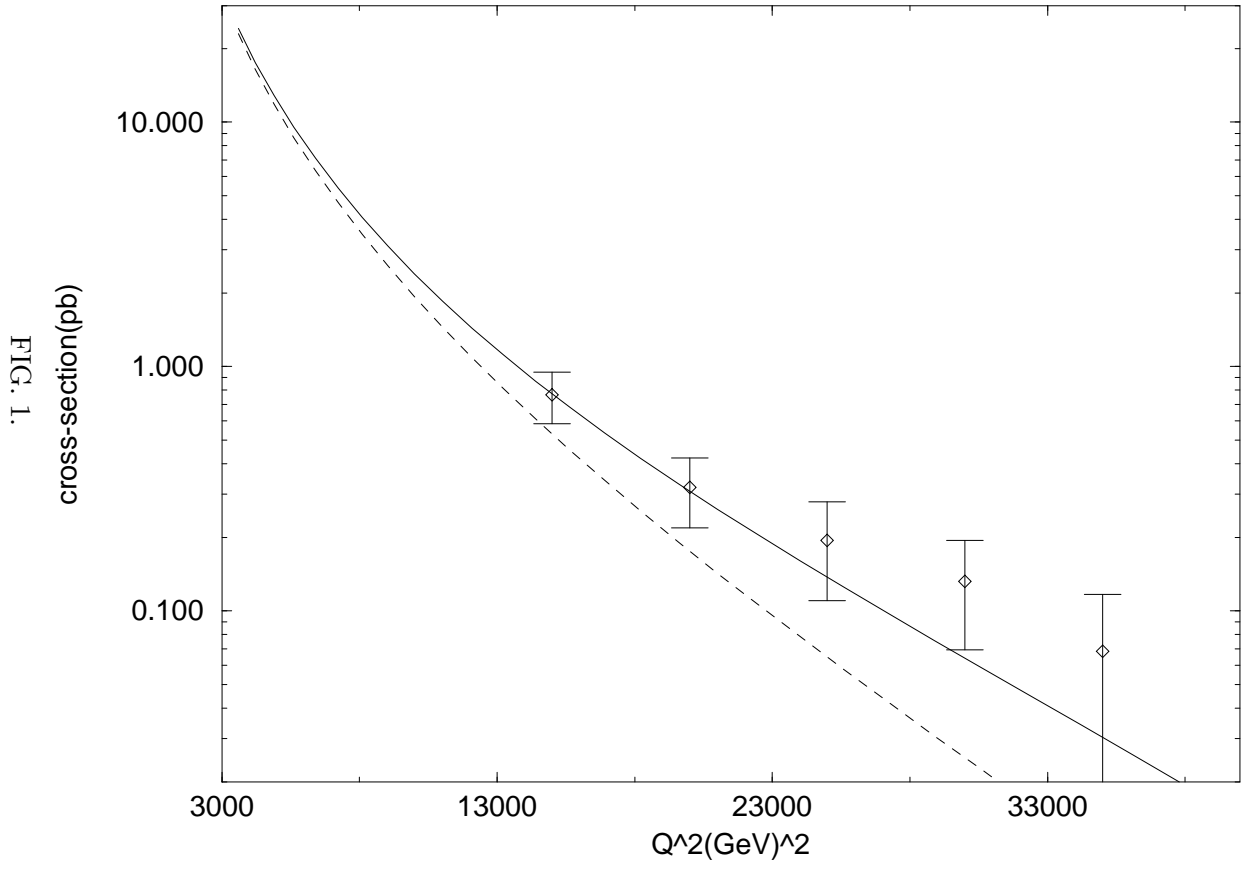
\begin{figure}
\centerline{ \DESepsf(herafig1.epsf width 12 cm) }
\bigskip
\caption {}
\vspace{3 cm}
\end{figure}
\begin{figure}
\centerline{ \DESepsf(herafig23.epsf width 12 cm) }
\smallskip
\vspace{3 cm}
\end{figure}

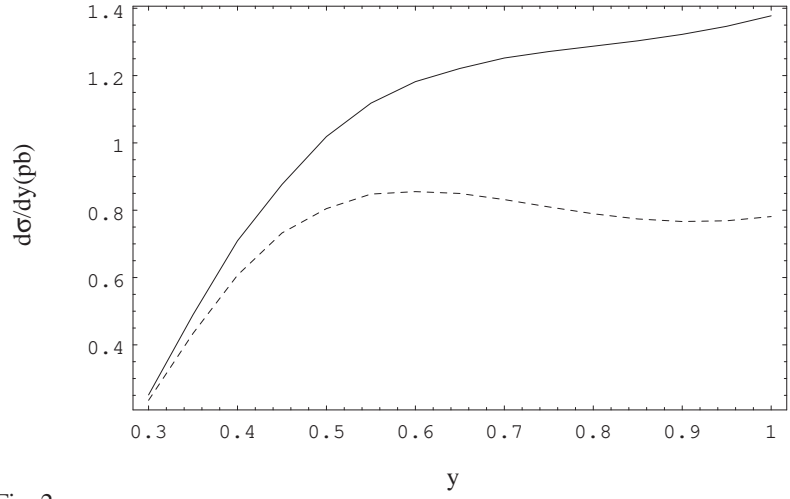
\end{document}